\documentclass{iau} 
\usepackage{graphicx}

\title[JD 11.~~Herschel PSC] 
{The Herschel\thanks{{\it Herschel} is an ESA space observatory with science instruments provided by European-led Principal Investigator consortia and with important participation from NASA.} Point Source Catalogue}

\author[G\'abor Marton et al.]   
{G. Marton$^1$, B. Schulz$^2$, B. Altieri$^3$, L. Calzoletti$^3$, Cs. Kiss$^1$, T. Lim$^3$, N. Lu$^2$, R. Paladini$^2$, 
A. Papageorgiou$^4$, C. Pearson$^5$, J. Rector$^2$, D. Shupe$^2$, I. Valtchanov$^3$, E. Vereb\'elyi$^1$, K. Xu$^2$}

\affiliation{$^1$Konkoly Observatory, Research Centre for Astronomy and Earth Sciences, Hungarian Academy of Sciences, Konkoly Thege Mikl\'os \'ut 15-17, 1121 Budapest, Hungary \\ email: {\tt marton.gabor@csfk.mta.hu} \\
$^2$NASA Herschel Science Center, California Institute of Technology, \\1200, East California Boulevard, Pasadena, CA 91125, USA \\ email: {\tt bschulz@ipac.caltech.edu} \\
$^3$European Space Astronomy Centre (ESAC)/ESA, \\PO Box 78, 28690 Villanueva de la Ca\~nada, Madrid, Spain \\
$^4$School of Physics and Astronomy, Cardiff University, The Parade, Cardiff CF24 3AA, UK \\
$^5$RAL Space, STFC Rutherford Appleton Laboratory, Didcot, Oxon, OX11 0QX, \\UK The Open University, Milton Keynes MK7 6AA, UK}

\pubyear{2015}
\volume{315}  
\jname{From interstellar clouds to star-forming galaxies: universal processes?}
\editors{A.C. Editor, B.D. Editor \& C.E. Editor, eds.}
\begin{document}

\maketitle

\begin{abstract}
The Herschel Space Observatory was the fourth cornerstone mission in the European Space Agency (ESA) science programme with excellent broad band imaging capabilities in the sub-mm and far-infrared part of the spectrum. Although the spacecraft finished its observations in 2013, it left a large legacy dataset that is far from having been fully scrutinised and still has a large potential for new scientific discoveries. This is specifically true for the photometric observations of the PACS and SPIRE instruments. Some source catalogues have already been produced by individual observing programs, but there are many observations that risk to remain unexplored. To maximise the science return of the SPIRE and PACS data sets, we are in the process of building the Herschel Point Source Catalogue (HPSC) from all primary and parallel mode observations. Our homogeneous source extraction enables a systematic and unbiased comparison of sensitivity across the different Herschel fields that single programs will generally not be able to provide. The catalogue will be made available online through archives like the Herschel Science Archive (HSA), the Infrared Science Archive (IRSA), and the Strasbourg Astronomical Data Center (CDS).

\end{abstract}

\firstsection 
\section{Introduction}

The Herschel Space Observatory (\cite{pilbratt2010}) was actively observing the sky from June, 2009 to April, 2013. Two of the three instruments on-board had deep broad-band imaging capabilities. The PACS instrument (\cite{poglitsch2010}) observed the sky at far-infrared wavelengths (70, 100 \& 160 $\mu$m), while the SPIRE instrument made observations in the sub-mm wavelength regime (250, 350 \& 500 $\mu$m). With a primary mirror of 3.5 meters in diameter, the angular resolution was far better than any space telescope before at these wavelengths.

During its lifetime, Herschel observed $\sim$10\% of the sky. Being mainly sensitive to cold dust, the photometric parts of the Guaranteed and Open Time Key Programs were geared towards distant and nearby dusty galaxies, the Galactic cirrus, protostars and pre-stellar cores, and distant Solar System Objects, all being key to understanding the star and galaxy formation history.

\section{The Catalogue}
The primary goal of our work is to compile an independent source table for each band. Band-merging will not be provided, as it appears only to be feasible with additional data and substantially more effort. 

\begin{figure}[b]
\begin{center}
 \includegraphics[angle=270,width=3.2in,trim=8cm 5cm 6cm 3cm]{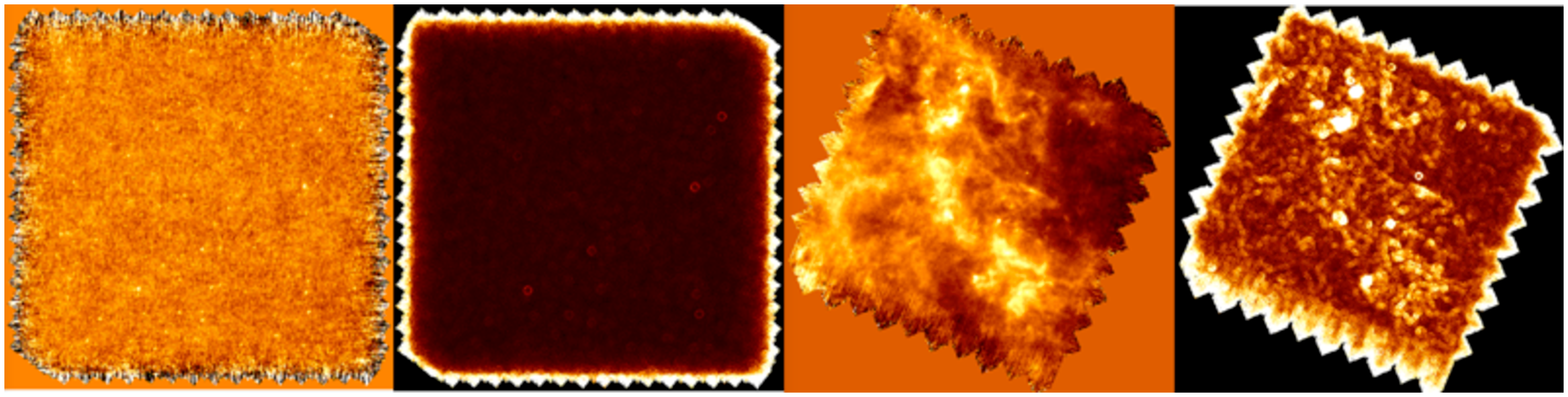} 
 \caption{Example maps and derived structure noise maps for PACS in the 160 $\mu$m filter band. First and second panels from the left show the Lockman Hole and its structure noise map, respectively. The third and fourth panels show field G334.65+2.67 and its structure noise map.}
   \label{fig1}
\end{center}
\end{figure}

Quality assessment in the far-infrared and sub-mm is not an easy task. The many features of the ISM seen at these wavelengths make the detection and photometry of point sources a challenging problem. Artificial sources were added to maps with different levels of complexity to test the completeness and photometric accuracy in various environments. Our results clearly showed that completeness and reliability are highly dependent on the brightness variation of the environment, and our detections need to be treated with special attention. To this end, we use structure noise (\cite{kiss2005}) as a metric to estimate the emission fluctuations around a given point in the sky which, in turn, provides a description of the noise at the location of each detected source. The structure noise contains both the celestial background variations and the instrumental noise, and it is used to derive detection uncertainties. Examples of structure noise maps are shown in Fig. \ref{fig1}.




To allow the statistical exploitation of the catalogue content, we will provide numerous flags indicating, e.g., if a source is extended, affected by a Solar System Object, variable, or located at the edge of the map.



The first release of the HPSC will be in early 2016. This work is not only yielding additional products to the scientific community, but has also helped the instrument teams to substantially improve the standard data processing and the final legacy products of the mission. Our homogeneously produced catalogue will help users to obtain information on millions of sources that were never before observed at far-IR and sub-mm wavelengths, and will allow for a better distinction of different source classes. Large scale studies, like on star formation rate and clustering will also benefit. Last but not least it will provide an excellent path finder for facilities like ALMA and SOFIA. 


\end{document}